\documentclass[twocolumn]{aastex62}
\usepackage{graphicx}
\usepackage{threeparttable, tablefootnote}
\usepackage{mhchem}
\usepackage{gensymb}

\received{June 14, 2019}
\revised{July 19, 2019}
\accepted{July 27, 2019}

\begin{document}

\title{\uppercase{The First Detection of \ce{^{13}C^{17}O} in a Protoplanetary Disk: \\ a Robust Tracer of Disk Gas Mass}}

\author[0000-0003-2014-2121]{Alice S. Booth}
\affiliation{School of Physics and Astronomy, University of Leeds, Leeds LS2 9JT, UK}
\email{pyasb@leeds.ac.uk}

\author[0000-0001-6078-786X]{Catherine Walsh}
\affiliation{School of Physics and Astronomy, University of Leeds, Leeds LS2 9JT, UK}

\author[0000-0003-1008-1142]{John D. Ilee}
\affiliation{School of Physics and Astronomy, University of Leeds, Leeds LS2 9JT, UK}

\author[0000-0003-2493-912X]{Shota Notsu}
\affiliation{Department of Astronomy, Graduate School of Science, Kyoto University, Kitashirakawa-Oiwake-cho, Sakyo-ku, Kyoto 606-8502, Japan}
\affiliation{ Leiden Observatory, Leiden University,
P.O. Box 9513, NL-2300 RA, Leiden, the Netherlands}

\author[0000-0001-8642-1786]{Chunhua Qi}
\affiliation{Harvard-Smithsonian Center for Astrophysics, Cambridge, MA 02138, USA}

\author[0000-0002-7058-7682]{Hideko Nomura}
\affiliation{Department of Earth and Planetary Science, Tokyo Institute of Technology, 2-12-1 Ookayama, Meguro-ku, Tokyo 152-8551, Japan}
\affiliation{National Astronomical Observatory Japan (NAOJ), Osawa 2-21-1, Mitaka, Tokyo 181-8588, Japan}

\author[0000-0002-5082-8880]{Eiji Akiyama}
\affiliation{Institute for the Advancement of Higher Education, Hokkaido University, Kita 17, Nishi 8, Kita-ku, Sapporo, Hokkaido 060-0817, Japan}

\begin{abstract}
Measurements of the gas mass are necessary to determine the planet formation potential of protoplanetary disks.  Observations of rare CO isotopologues are typically used to determine disk gas masses; however, if the line emission is optically thick this will result in an underestimated disk mass. With ALMA we have detected the rarest stable CO isotopologue, \ce{^{13}C^{17}O}, in a protoplanetary disk for the first time. We compare our observations with the existing detections of \ce{^{12}CO}, \ce{^{13}CO}, \ce{C^{18}O} and \ce{C^{17}O} in the HD~163296 disk. Radiative transfer modelling using a previously benchmarked model, and assuming interstellar isotopic abundances, significantly underestimates the integrated intensity of the \ce{^{13}C^{17}O} J=3-2 line. Reconciliation between the observations and the model requires a global increase in CO gas mass by a factor of 3.5.
This is a factor of 2-6 larger than previous gas mass estimates using \ce{C^{18}O}. We find that \ce{C^{18}O} emission is optically thick within the snow line, while the \ce{^{13}C^{17}O} emission is optically thin and is thus a robust tracer of the bulk disk CO gas mass.
\end{abstract}



\section{Introduction}
The mass of a disk sets a limit on the material available for forming a planetary system and can influence the mode of giant planet formation.   
Most disk gas masses rely on observations of CO that are extrapolated to a total gas mass by assuming a constant \ce{CO}/\ce{H_2} abundance ratio in the disk.  However, if the line emission is optically thick this will result in an underestimated disk mass \citep[e.g.,][]{2017ASSL..445....1B}. Spatially resolved \ce{^{13}C^{16}O} and \ce{^{12}C^{18}O} line emission are now commonly detected in protoplanetary disks and have been used to determine disk gas masses \citep[e.g.,][]{2016ApJ...828...46A, 2017ApJ...844...99L}.
The second-most rarest CO isotopologue, \ce{^{13}C^{18}O}, has been detected in the TW Hya disk and this emission has been proposed to be optically thin and traces the disk midplane whereas the  \ce{C^{18}O} emission is optically thick within the midplane snowline \citep{2017NatAs...1E.130Z}. The under-estimation of disk gas mass due to optically thick emission will be more significant in more massive gas-rich disks, i.e., those around Herbig Ae/Be stars versus those around T Tauri stars.

Low CO gas masses, with respect to the dust mass, have been consistently measured in disks and complementary HD observations imply that this is because of the depletion of gas-phase CO in disks relative to that in the ISM \citep{2013Natur.493..644B, 2016ApJ...831..167M}. CO can be depleted from the gas phase via freeze out onto the icy grains in the cold midplane, and subsequent conversion to \ce{CO_2} and more complex organic species e.g. \ce{CH_3OH}  \citep[e.g.][]{2018arXiv180802220B}. Additionally, photodissociation via far-UV radiation destroys CO in the upper disk atmosphere, and 
isotope selective photodissociation can enhance the various isotopologue ratios relative to \ce{^{12}C^{16}O} in the atmosphere \citep[e.g.][]{2014A&A...572A..96M}. These chemical effects make the conversion from CO gas mass to total gas mass a non-trivial task.

The protoplanetary disk around HD~163296  
has been well characterised with the Atacama Large Millimeter/submillimeter Array (ALMA). Band 6 and 7 observations show rings in both the continuum and the CO gas emission \citep[e.g.][and see Figure 1a]{2016PhRvL.117y1101I, 2019ApJ...875...96N}. There are four proposed $\approx$~0.5~-~2$~M_{J}$ planets in this disk inferred from the dust and gas rings, and deviations from Keplerian motion in the CO gas kinematics \citep{2016PhRvL.117y1101I, 2018ApJ...860L..13P, 2018ApJ...860L..12T, 2018ApJ...857...87L}. Recently, $\approx$~5~au resolution observations of the continuum emission revealed an additional gap and ring in the inner disk as well as an azimuthal asymmetry in one of the previously detected rings \citep[see Figure 1a,][]{2018ApJ...869L..49I}. Hence, the proposed planet-induced structures in the HD~163296 disk make it an excellent observational laboratory to study planet formation.

We present the first detection of \ce{^{13}C^{17}O} in a protoplanetary disk providing a strong constraint on the CO gas mass in the HD~163296 disk.

\section{Observations}

HD~163296 was observed with ALMA in Band 7 during Cycle 3 on 2016 September 16 (2015.1.01259.S, PI: S.~Notsu). See \citet{2019ApJ...875...96N} for details on the calibration and self-calibration of the data. The spectral windows have a resolution of 1953.125~kHz.  There are 14 \ce{^{13}C^{17}O} J=3-2 hyperfine structure lines that lie between 321.851 and 321.852~GHz. All lines lie within a frequency range less than the spectral resolution of the data; hence, we are observing the blending of all of the lines. The \ce{^{13}C^{17}O} molecular data we use is from \citet{Klapper_2003} and was accessed via the Cologne Database for Molecular Spectroscopy \citep[CDMS,][]{2005JMoSt.742..215M}. We detected \ce{^{13}C^{17}O} initially via a matched filter analysis\footnote{A python-based open-source implementation of VISIBLE is available at \url{http://github.com/AstroChem/VISIBLE}} \citep{2018AJ....155..182L} using a Keplerian mask assuming a disk position angle of 132\degree~and an inclination of 42\degree~\citep[e.g.][]{2016PhRvL.117y1101I}.
The resulting S/N ratio is $\approx 3.5$. The filter response is shown in Figure~1b with the black line marking the \ce{^{13}C^{17}O} J=3-2 transition after correction for the source velocity (5.8~$\mathrm{km~s^{-1}}$). 

The line imaging was conducted using CLEAN with CASA version 4.6.0. The native spectral resolution of the data is 1.8~$\mathrm{km~s^{-1}}$; however, in order to optimise the S/N the final images were generated with a 3~$\mathrm{km~s^{-1}}$ channel width and a $uv$ taper of 0\farcs5 resulting in a synthesised beam of 0\farcs87 x 0\farcs51~(100\degree). 
Figures 1c and 1d present the \ce{^{13}C^{17}O} integrated intensity map and the intensity-weighted velocity map, respectively. The integrated intensity map was made using channels $\pm~6~\mathrm{km~s^{-1}}$ about the source velocity. The peak integrated intensity is 0.55~$\mathrm{Jy~beam^{-1}~km~s^{-1}}$ with an rms noise level of 0.08~$\mathrm{Jy~beam^{-1}~km~s^{-1}}$ (S/N~=~7), that was extracted from the spatial region beyond the detected line emission. The intensity-weighted velocity map was made in the same manner but also with a 3$\sigma$ clip.

We also use archival data to benchmark our modelling, 
including the \ce{^{12}C^{16}O}, \ce{^{13}C^{16}O}, \ce{^{12}C^{18}O} J=2-1 transitions observed from \citet{2016PhRvL.117y1101I}, and the \ce{^{12}C^{16}O} J=3-2 ALMA Science Verification data.\footnote{\url{https://almascience.nrao.edu/alma-data/science-verification}} 
All integrated intensity maps were de-projected and azimuthally averaged and are shown in Figures 3a to 3e. 
The errors are the standard deviation of intensity of the pixels in each bin divided by the number of beams per annulus \citep[e.g.][]{2018A&A...614A.106C}. 
In addition, Figure 3e shows the \ce{^{12}C^{17}O} J=3-2 total integrated intensity value with its associated errors \citep{2011ApJ...740...84Q}.
All data plotted assumes a source distance of 122~pc \citep{1998A&A...330..145V}.
Although {\sc gaia} DR2 puts this source at 101.5~pc \citep{2018A&A...616A...1G}, in order to compare to previous analyses we use the previous value. We discuss the impact of the revised distance in Section 4.

\begin{figure*}
\centering
\includegraphics[width=\hsize]{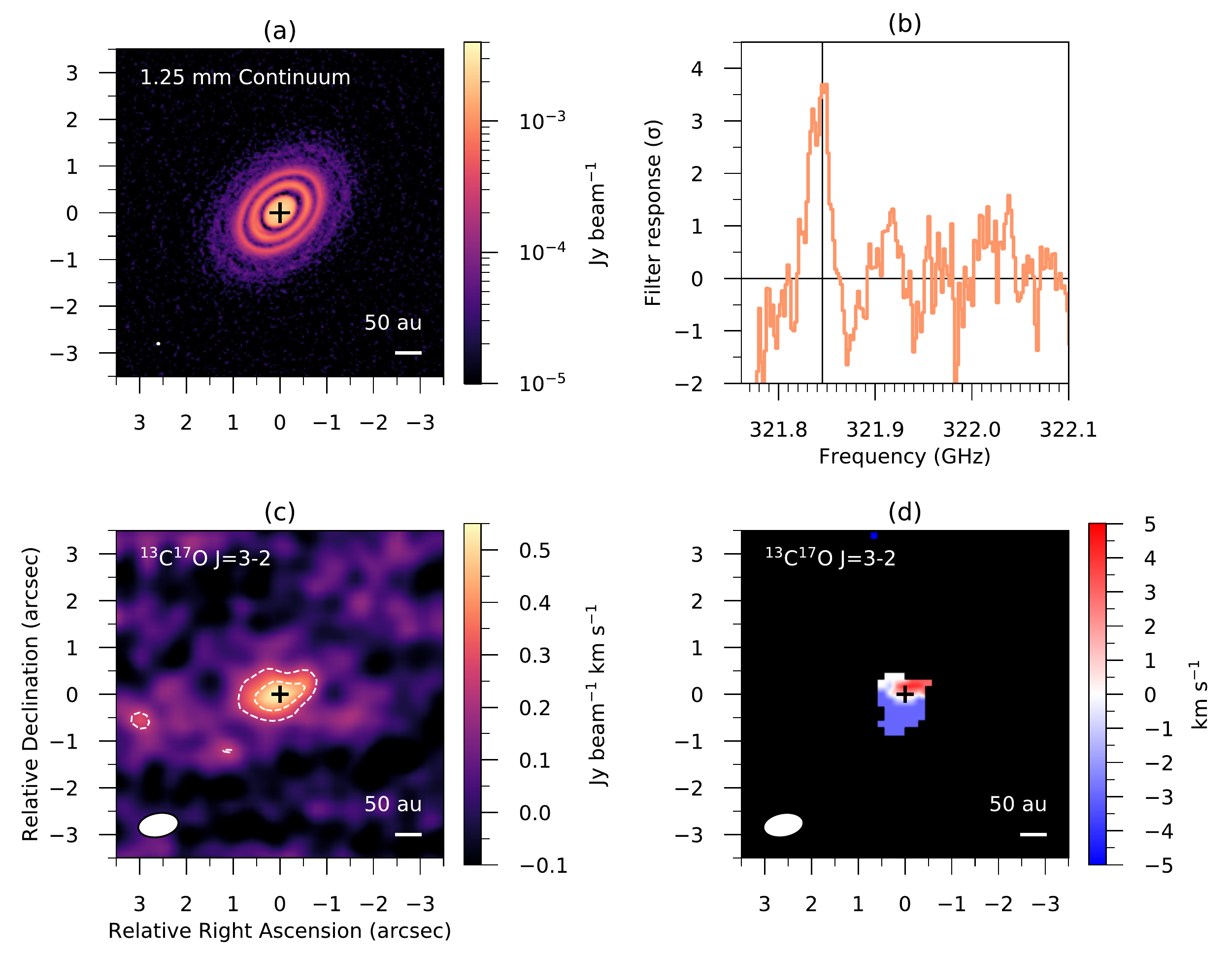}
\caption{
a) The 1.25~mm continuum image from \citet{2018ApJ...869L..49I}.
b) The matched filter response for the \ce{^{13}C^{17}O} J=3-2 detection where the black line marks the frequency of the hyper-fine transitions.
c) The \ce{^{13}C^{17}O} J=3-2 integrated intensity map where the white dashed contours mark 3 and 5 $\sigma$.
d) The \ce{^{13}C^{17}O} J=3-2 intensity-weighted velocity map.}
\end{figure*}

\section{Analysis}
Previous observations of the HD~163296 disk with the SMA and ALMA have detected multiple \ce{CO} isotopologues: \ce{^{12}C^{16}O}, \ce{^{13}C^{16}O},
\ce{^{12}C^{18}O} and \ce{^{12}C^{17}O} \citep{2011ApJ...740...84Q, 2016PhRvL.117y1101I}.
The models that were used to reproduce the line emission in \citet{2011ApJ...740...84Q}  
recover the following global isotope ratios;
\begin{gather*}
n(\ce{^{12}C^{16}O})/n(\ce{^{13}C^{16}O}) = 67~\pm~8,\\
n(\ce{^{12}C^{16}O})/n(\ce{^{12}C^{18}O}) = 444~\pm~88,\\
n(\ce{^{12}C^{18}O})/n(\ce{^{12}C^{17}O}) = 3.8~\pm~1.7,
\end{gather*}
where $n(\ce{^{X}C^{Y}O})$ is the number density of the molecule.
This is consistent with the carbon and oxygen isotope ratios observed in the ISM \citep{0034-4885-62-2-002}. 

We make a first estimate of the column density of gas traced by the 
\ce{^{13}C^{17}O} emission under the assumption of optically thin emission in local thermodynamic equilibrium. Following \citealt{2019A&A...623A.124C} (their Equation 1, with molecular data obtained from CDMS, \citealt{2005JMoSt.742..215M}), the average column density for the \ce{^{13}C^{17}O} within 50~au, assuming an excitation temperature of 50~K, is $\mathrm{7.1~\times~10^{15}\mathrm{~cm^{-2}}}$.
This is equivalent to a $n_{\rm H}$ column density of
$\mathrm{2.65~\times~10^{25}\mathrm{~cm^{-2}}}$ ($\mathrm{44.4~g~cm^{-2}}$) at 50~au. 
In comparison, the corresponding value for the \ce{^{12}C^{18}O}
is $\mathrm{1.7~\times~10^{16}~\mathrm{cm^{-2}}}$, resulting in a 
$n(\ce{^{12}C^{18}O})$/$n(\ce{^{13}C^{17}O}$) ratio of 2.5.
Under the assumption that both the lines are optically thin (and taking the previously derived isotopic ratios), this value is a factor of 100 too small. Therefore, the \ce{^{12}C^{18}O} line emission is optically thick 
and the resulting gas mass derived from this tracer will be underestimated.

%
%
%
%
%
To quantify this more robustly, 
we utilise an existing disk model that has been shown
to fit emission lines from multiple CO isotopologues (\ce{^{12}C^{16}O}, \ce{^{13}C^{16}O}, \ce{^{12}C^{18}O} and \ce{^{12}C^{17}O})
to model our \ce{^{13}C^{17}O} detection \citep{2011ApJ...740...84Q}.
The density (hydrogen nuclei density, $n_{\rm H}$) and temperature of the disk are shown in Figures 2a and 2b. 
The CO abundance distribution, shown in Figure 2c, was determined by setting $n(\ce{CO})$ to a constant fractional abundance of $6.0\times10^{-5}$ with respect to \ce{H2} in the molecular layer following \citet{2011ApJ...740...84Q}. This assumes that 86\% of volatile carbon inherited by the disk is in the form of CO \citep{1982ApJS...48..321G}.  This abundance was reduced by a factor of $10^{-4}$ in the midplane where $\mathrm{T_{gas}~\leq~19~K}$ and by a factor of $10^{-8}$ in the atmosphere where the vertically-integrated hydrogen column density, $\sigma(n_{\rm H})$, from the disk surface is $ < 1.256~\times~10^{21}~\mathrm{cm^{-2}}$.
The depleted value in the midplane is consistent with the CO abundances derived from chemical models including non-thermal desorption \citep{2010ApJ...722.1607W}. The photodissociation and freeze out boundaries are shown in white contours overlaid on Figures 2a and 2b.

\begin{figure*}
\centering
\includegraphics[width=\hsize]{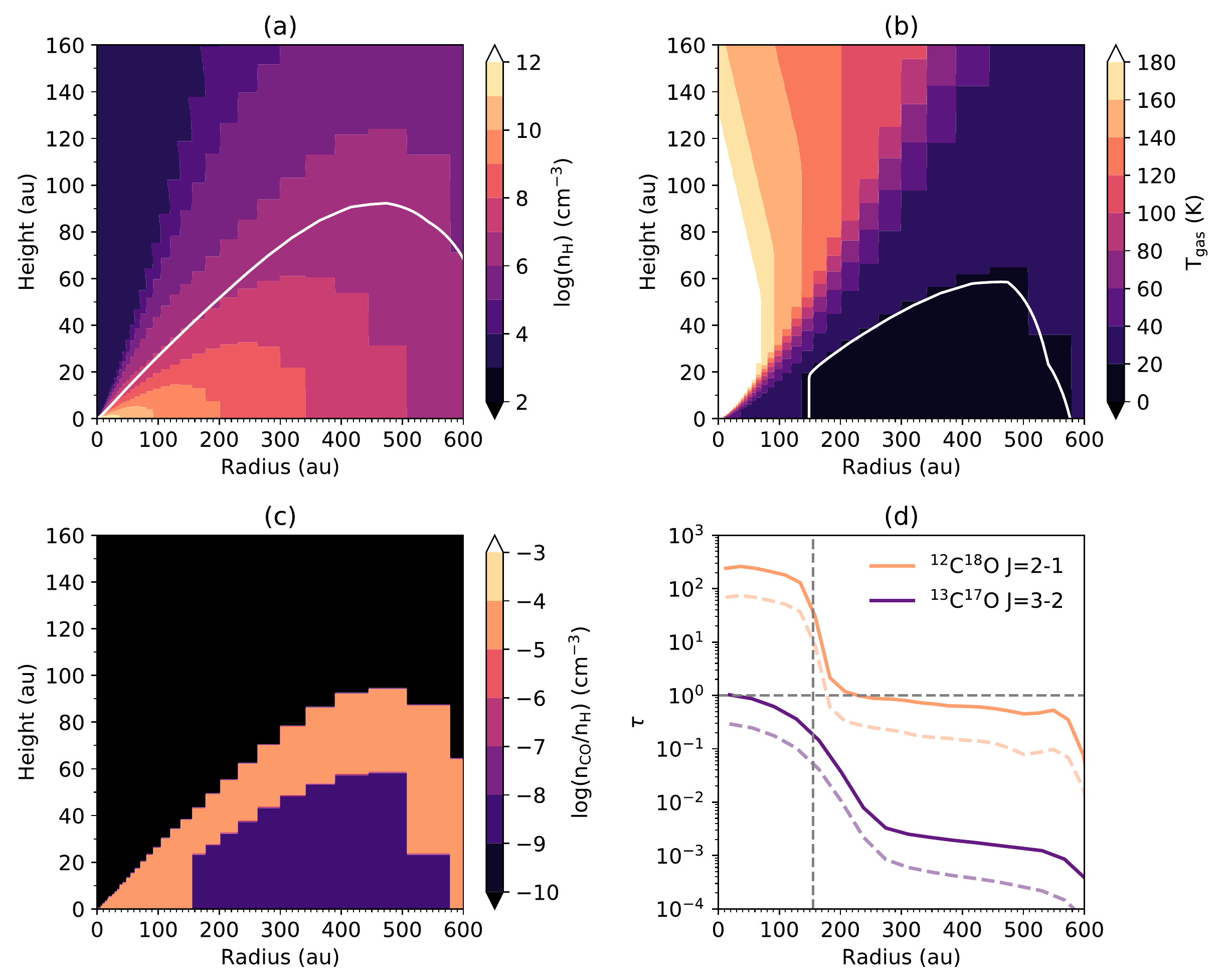}
\caption{The disk physical structure from \citet{2011ApJ...740...84Q}. 
a) The $n_{\rm H}$ density.
b) The gas temperature.
The white contours mark $\Sigma(n_{\rm H})$~=~$1.256~\times~10^{21}~{\rm cm}^{-2}$ and 
$T_{\rm gas}~=~19$K, respectively.
c) The $n(\ce{CO})/n_{\rm H}$ distribution.
d) The radially averaged optical depth ($\mathrm{\tau}$) of the \ce{^{12}C^{18}O} J=2-1 and \ce{^{13}C^{17}O} J=3-2 transitions from Model 1 (light purple and orange dashed lines) and Model 2 (dark purple and orange solid lines) assuming a face on disk. The vertical dashed line marks the location of the CO snowline in both
models (155~au) and the horizontal dashed line marks where $\mathrm{\tau}=1$.}
\end{figure*}

The first model that we test, Model 1, uses a constant \ce{^{13}C^{17}O} fractional abundance  of $5.39~\times~10^{-10}$ relative to \ce{H2}. This assumes isotope ratios that are consistent with the observations and modelling from \citet{2011ApJ...740...84Q}. 
Model 1 has a total disk mass of 0.089~M$_{\odot}$.
Using the CDMS data for \ce{^{13}C^{17}O} we generated a LAMDA-like file in order to model the J=3-2 hyper-fine components in LIME\footnote{\url{https://github.com/lime-rt/lime}} \citep[the Line Modeling Engine,][]{2010A&A...523A..25B}. Synthetic images cubes were computed assuming the appropriate position angle and inclination of the source, and the resulting images were smoothed with a Gaussian beam to the spatial resolution of the observations using the CASA task, \textsc{imsmooth}. The generated integrated intensity map was then de-projected and azimuthally averaged. The radial profiles from Model 1 (orange) are shown alongside the observations in Figures 3a to 3e. 

Model 1 under-predicts the \ce{^{13}C^{17}O} peak emission in the integrated intensity map by a factor of 2.5, yet provides a reasonable fit to the other lines (within a factor of two). The higher spatial resolution observations are effected by dust opacity within $\approx~50$~au \citep[see][]{2016PhRvL.117y1101I}; therefore we focus on reproducing the data beyond 50~au.

\begin{figure*}[p!]
\centering
\includegraphics[width=\hsize]{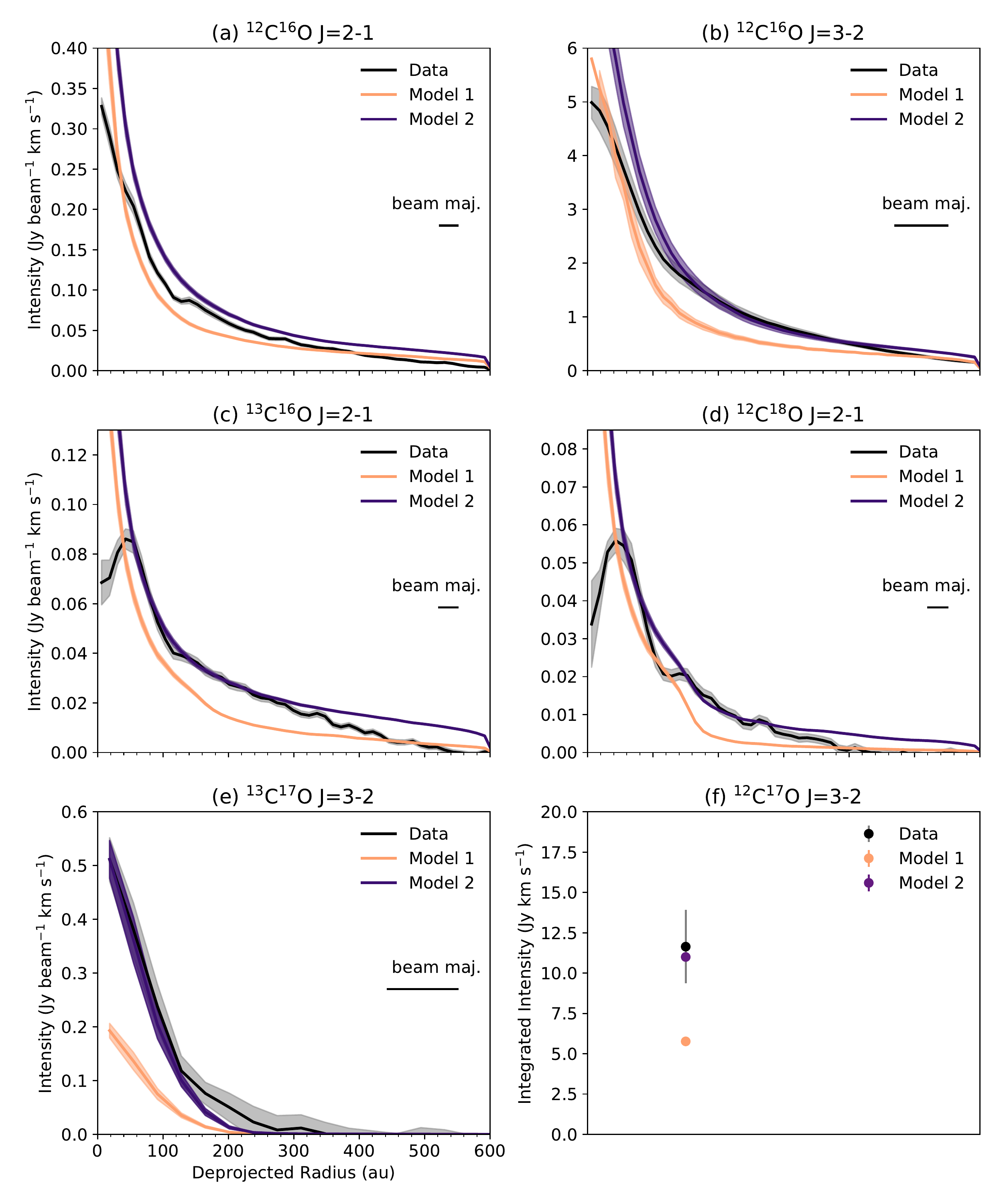}
\caption{a-e) The de-projected and azimuthally averaged radial profiles of the observed and modelled CO lines. 
f) The value of integrated flux of the observed and modelled \ce{C^{17}O} J=3-2 line.  
The shaded regions are the errors as described in the text.
Model 1 has a total disk mass of 0.089~M$_{\odot}$ and
Model 2 has a total disk mass of 0.31~M$_{\odot}$.}
\end{figure*}

To better fit the \ce{^{13}C^{17}O} observations we globally increase the gas mass of Model 1. This was done by initially multiplying $n_{\rm H}$ by a factor of 1.5 and then increasing this factor in steps of 0.5 until the best by-eye fit of 3.5 was found. The results for Model 2 are shown in Figure 3 (purple). Model 2 provides a better fit to all of the lines. This model assumes a smooth radial gas density structure contrary to the most recent observations.
However, our work is focused on reproducing the global disk mass rather than the underlying small scale gas surface density variations. Model 2 has a total disk mass of 0.31~M$_{\odot}$.

We note that a similar fit can be obtained using a different CO snowline location at 90 au as determined in \citet{2015ApJ...813..128Q}. This requires a corresponding increase in gas mass ($\times$~3.5) within the snowline, and we obtain a similar \ce{^{12}C^{18}O} column density profile as in \citet{2015ApJ...813..128Q} beyond the snowline.
Both of these models use the same underlying physical structure but have different CO snowline locations and levels of CO depletion beyond the snowline. The \citet{2011ApJ...740...84Q} model has simpler assumptions regarding the freeze-out of CO, consistent with other work \citep[e.g.,][]{2014ApJ...788...59W}, and was found to be a slightly better fit to the observations.

We also generated optical depth maps for a face on disk to recover the maximum value of $\mathrm{\tau}$ for each transition. We then radially averaged these maps and plot the resulting optical depth of the \ce{^{12}C^{18}O} J=2-1 and \ce{^{13}C^{17}O} J=3-2 transitions for both models in Figure 2d.  It can be seen that \ce{^{12}C^{18}O} is optically thick within the CO snowline (155~au) in both models, whereas the \ce{^{13}C^{17}O} remains optically thin across the full radial extent of the disk.

\section{Discussion}

\subsection{Comparison to other mass estimates}

Using observations of \ce{^{13}C^{17}O} we derive a new gas mass for the HD~163296 disk of 0.31~M$_{\odot}$.
The total disk mass depends on the gas to dust mass ratio (g/d), and using the dust mass from \citealt{2007A&A...469..213I} we find a g/d $\approx$ 260.  Here we compare our results to previous works.

The HD~163296 disk has been well studied and there are many mass measurements in the literature. In general, our estimate is the highest by a factor of 2 to 6 compared to previous studies using \ce{^{12}C^{18}O} (e.g. 0.17~M$_{\odot}$ and 0.048~M$_{\odot}$ from \citealt{2007A&A...469..213I} and \citealt{2016ApJ...830...32W} respectively). There are a range of g/d values in the literature that span four orders of magnitude. \citet{2012A&A...538A..20T} and \citet{2016MNRAS.461..385B} models require a low g/d~$=~20$. \citet{2016PhRvL.117y1101I} have a radially varying g/d covering a range from $\approx$~30 to $\approx$~1100. 
Recent work from \citet{2019ApJ...878..116P} recover a total disk mass of 0.21~M$_{\odot}$ with a high g/d $\sim 10^{4}$ in the outer disk. The one documented mass higher than our result is 0.58~M$_{\odot}$ with a g/d~$=~$350 \citep{2019PASP..131f4301W}.
The inconsistencies in these mass measurements and g/d from different models may be explained by trying to recover the gas density structure with optically thick lines. CO remains the best and most accessible tracer of mass that we have for disks \citep{2017ApJ...849..130M}, but robust lower limits to the gas mass can only be made by targeting the most optically thin isotopologues (\ce{^{12}C^{17}O}, \ce{^{13}C^{18}O}, and \ce{^{13}C^{17}O}). 

These masses have all been determined using a source distance of 122~pc. Considering the revised distance of 101.5~pc, the total disk gas mass from our work is thus 0.21~$M_{\odot}$ ($\mathrm{mass~\propto~flux~\propto~distance^2}$).

\subsection{The impact of CO chemistry on the disk mass}

CO is susceptible to isotope-selective photodissociation which can reduce the abundance of the rarer isotopologues relative to \ce{^{12}C^{16}O} in the disk atmosphere. We find that the observations are well fit with interstellar isotopic abundances. Because the \ce{^{12}C^{16}O}, \ce{^{13}C^{16}O} and \ce{^{12}C^{18}O} line emission is optically thick, testing the significance of isotope-selective photodissociation in this disk requires higher sensitivity observations of the rarer isotopologues.

Observations have shown that CO is depleted with respect to \ce{H_2} in disks; however, without a better tracer of the \ce{H2} column density, e.g., HD, the level of depletion is difficult to constrain. Carbon depletion effects are less significant in warmer disks around Herbig Ae stars compared to their T Tauri counterparts. Observations show moderate carbon depletion in the Herbig disk around HD~100546 with a model-derived [C]/[H] abundance ratio of 0.1 to 1.5~$\times$~$10^{-4}$ \citep{2016A&A...592A..83K}, and the value for CO adopted in our model is within this range. Consistent with this, models have also suggested that these disks have a close to canonical $n(\ce{CO})/n(\ce{H2})$ abundance \citep{2018arXiv180802220B}. 
These two chemical effects (isotope-selective photodissocation and carbon depletion) imply that our gas mass estimate is a lower limit. 

\subsection{Constraints on the location of the CO snowline}
Locating the midplane CO snowline in disks is difficult due to the high optical depth of the more abundant CO isotopologues and the vertical temperature gradient of the disk. The location of the CO snowline can be determined directly by observing less abundant, optically thin, CO isotopologues, or by detecting molecules that, due to chemistry, peak in abundance at a location related to the snowline. 
\citet{2011ApJ...740...84Q} use observations CO and put the snowline at 155~au at 19~K: follow up work suggested that the snowline was instead at 90~au and 25~K \citep{2015ApJ...813..128Q}.
Out of these two scenarios our \ce{^{13}C^{17}O} observations fit best with the former option given the data in hand. The relationship between the cations, \ce{DCO+} (ring from 110 and 160~au) and \ce{N_2H^+} (inner edge of the ring at 90~au), and the location of the 
midplane CO snowline is not trivial, but both species have been detected in this disk \citep{2013A&A...557A.132M, 2015ApJ...813..128Q}. 
Our analysis shows that the \ce{C^{18}O} emission in both models tested is optically thick, and thus cannot be used to easily locate the midplane CO snowline.
However, the \ce{^{13}C^{17}O} emission is optically thin, so future observations at a higher spatial resolution and sensitivity could be used to directly constrain the radius of the midplane CO snowline. 
The new source distance from {\sc gaia} puts the proposed snowline locations at 75~au and 128~au. The former location is close one of the observed dust gaps in the disk and it may be the case that the drop in CO surface density detected here is due to gas depletion rather than the snowline. It is important to note that the snowline is not a simple sharp transition at the condensation temperature, but is instead determined by the balance of the rates of freeze out and thermal desorption, which should be considered in future disk models.

\subsection{Is the disk gravitationally stable?}

The potential exoplanet population currently probed with ALMA, via the ringed depletion of continuum emission, are gas giant planets on wide orbits. In the case of HD~163296 this would imply a multiple giant planet system and indeed, the presence of such a system has already been proposed \citep{2016PhRvL.117y1101I, 2018ApJ...860L..13P, 2018ApJ...860L..12T, 2018ApJ...857...87L}. The formation of massive planets on wide orbits can in some cases be achieved by core accretion, but a more economical route might involve the gravitational fragmentation of the outer regions of the disk \citep{2011ApJ...731...74B}.  Our new, higher disk mass estimate prompts us to investigate whether such processes may have occurred (or be occurring) in the HD~163296 disk.

The stability of a disk against fragmentation can be quantified via the Toomre Q parameter \citep{1964ApJ...139.1217T}: $$ Q = \frac{c_s \kappa}{\pi G \Sigma} $$ where $c_s$ is the sound speed of the gas, $\kappa$ is the epicyclic frequency (equal to the angular velocity $\Omega$ in a Keplerian disk) and $\Sigma$ is the surface density of the gas. 
Toomre Q values of 1 or less imply that the disk is susceptible to fragmentation, but simulations have shown that disks with Q$\lesssim$1.7 begin to undergo instabilities in the form of non-axisymmetric spirals \citep{2007prpl.conf..607D}.
We calculate Q across the disk (Figure 4, orange) accounting for the lower mass due to the new source distance, assuming a g/d of 260, and the midplane temperature structure of our model (Figure 2b). We find that the minimum value of Q is $\approx$~6 at $\approx$~110~au, suggesting that the disk is currently gravitationally stable (in agreement with recent work from \citealt{2019ApJ...878..116P}).

\begin{figure}
    \centering
    \includegraphics[width=\hsize]{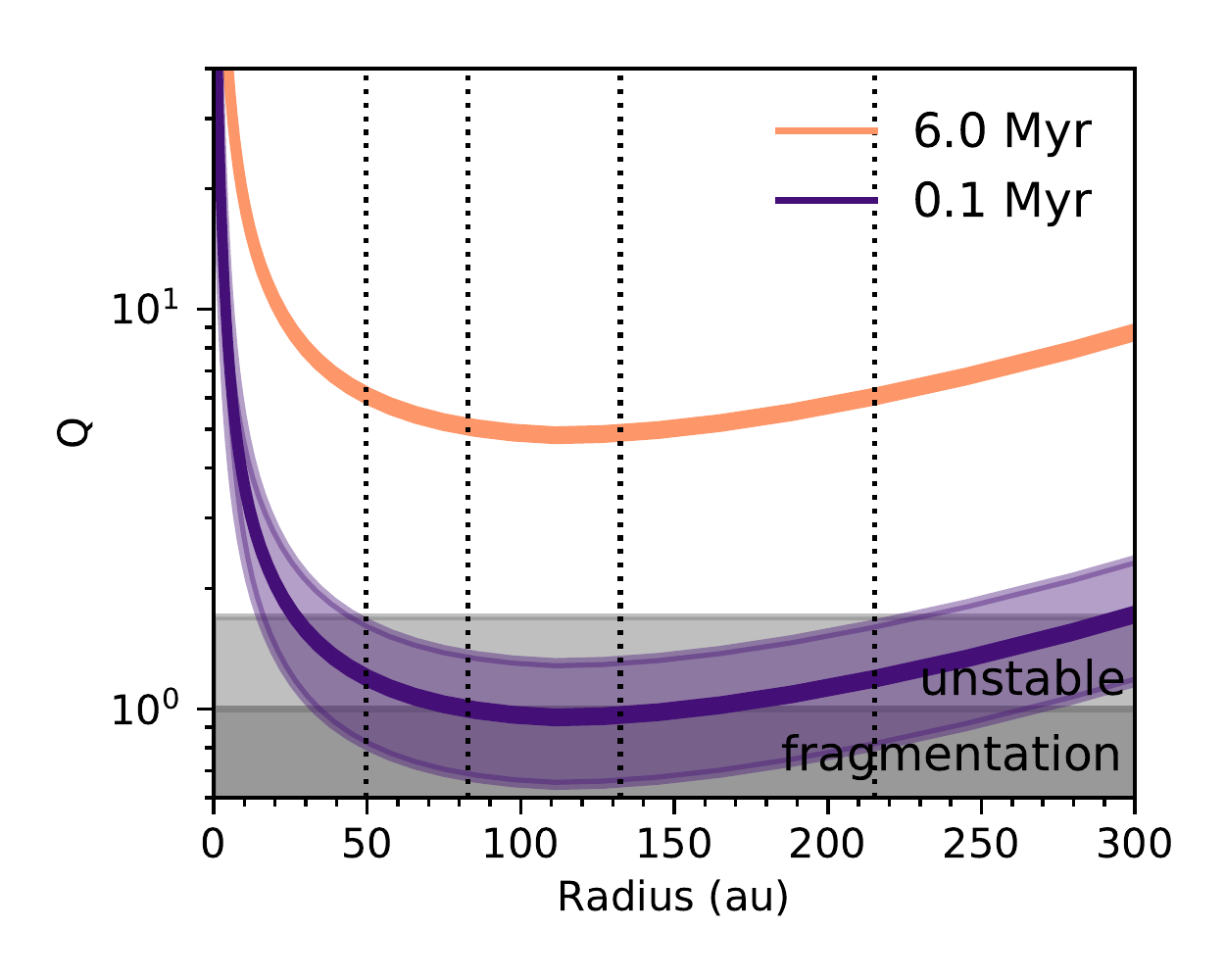}
    \caption{Toomre Q parameter for the new disk mass we derive in Model 2 and extrapolated back to 0.1~Myr. The shaded regions incorporate both the errors in stellar age and mass accretion rate.  Vertical dashed lines mark the radial positions of gaps in the mm dust and kinematic perturbations which may be due to protoplanets.}
    \label{fig:my_label}
\end{figure}

The relatively large age of the HD~163296 system 
brings into question its stability earlier in its lifetime.  The determination of previous disk masses is complicated by processes including episodic accretion \citep[e.g.][]{2013ApJ...776...44M} and the decrease in accretion rate with time \citep[e.g.][]{2014A&A...570A..82V}.  The magnitude of these effects are still under debate \citep[see][for a review]{2016ARA&A..54..135H}, so we therefore assume that all of the accreted mass once resided in the disk, and that the accretion rate ($\dot{M}$) has been constant over disk lifetime. HD~163296 has an estimated age of $6.03^{+0.43}_{-0.14}$~Myr and $\log \dot{M} = -6.81^{+0.16}_{-0.15}$\,M$_{\odot}$\,yr$^{-1}$ (Wichittanakom et al., in prep.); thus under these assumptions, we estimate the disk mass at 0.1~Myr to be $1.13_{-0.28}^{+0.51}$~$M_{\odot}$.  

The resulting minimum Toomre Q values for this star-disk configuration\footnote{We note this does not account for any change in the stellar mass over this time period, which would decrease Q further.} would be in the range of 1.3--0.7 (Figure 4, purple), placing regions of the disk from $\sim$50--220\,au in the regime of instability.  Such behaviour early in the disk lifetime has implications for the trapping and growth of dust \citep{2004MNRAS.355..543R} and the chemical composition of the disk \citep{2015MNRAS.453.1147E}. This previous unstable state could also be the source of the four massive planets currently proposed to reside in the disk around HD~163296.

\section{Conclusions}
We have presented the first detection of \ce{^{13}C^{17}O} in a protoplanetary disk showcasing the power of this optically thin isotopologue as a tracer of disk gas mass.
This work provides robust evidence that disks are more massive than previously assumed \citep[see also][]{2019arXiv190402127Z}.  Future observations of this tracer in more sources may help to address the discrepancy between the masses of disks and the observed exoplanet population \citep{2018A&A...618L...3M}.

\begin{acknowledgements}
We thank an anonymous referee for their constructive comments that improved the clarity of several sections of the paper. 
We thank Andrea Isella for the \ce{^{12}CO}, \ce{^{13}CO} and \ce{C^{18}O} J=2-1 data which was vital in our model comparison.
We also thank Chumpon Wichittanakom for providing us with updated stellar parameters for HD~163296 pre-publication of their paper. 
AB acknowledges the studentship funded by the Science and Technology Facilities Council of the United Kingdom (STFC). 
CW acknowledges funds from the University of Leeds.  
JDI and CW acknowledge support from the STFC under ST/R000549/1.
SN is grateful for support from JSPS (Japan Society for the Promotion of Science) Overseas Research Fellowships and the ALMA Japan Research Grant of NAOJ Chile Observatory, NAOJ-ALMA-211.
This work is supported by MEXT/JSPS KAKENHI Grant Numbers 16J06887, 17K05399, 19K03910.
Part of ALMA Data analysis was carried out on the Multi-wavelength Data Analysis System operated by the Astronomy Data Center (ADC), National Astronomical Observatory of Japan. 
This paper makes use of the following ALMA data: 
2011.0.00010.SV, 2013.1.00601.S and 2015.1.01259.S.
ALMA is a partnership of European Southern Observatory (ESO) (representing its member states), National Science Foundation (USA), and National Institutes of Natural Sciences (Japan), together with National Research Council (Canada), National Science Council and Academia Sinica Institute of Astronomy and Astrophysics (Taiwan), and Korea Astronomy and Space Science Institute (Korea), in cooperation with the Republic of Chile. The Joint ALMA Observatory is operated by ESO, Associated Universities, Inc/National Radio Astronomy Observatory (NRAO), and National Astronomical Observatory of Japan.


\end{acknowledgements}

\bibliographystyle{aasjournal}

\begin{thebibliography}{}
\expandafter\ifx\csname natexlab\endcsname\relax\def\natexlab#1{#1}\fi
\providecommand{\url}[1]{\href{#1}{#1}}

\bibitem[{{Ansdell} {et~al.}(2016){Ansdell}, {Williams}, {van der Marel},
  {Carpenter}, {Guidi}, {Hogerheijde}, {Mathews}, {Manara}, {Miotello},
  {Natta}, {Oliveira}, {Tazzari}, {Testi}, {van Dishoeck}, \& {van
  Terwisga}}]{2016ApJ...828...46A}
{Ansdell}, M., {Williams}, J.~P., {van der Marel}, N., {et~al.} 2016, \apj,
  828, 46

\bibitem[{{Bergin} \& {Williams}(2017)}]{2017ASSL..445....1B}
{Bergin}, E.~A., \& {Williams}, J.~P. 2017, in Astrophysics and Space Science
  Library, Vol. 445, Astrophysics and Space Science Library, ed. M.~{Pessah} \&
  O.~{Gressel}, 1

\bibitem[{{Bergin} {et~al.}(2013){Bergin}, {Cleeves}, {Gorti}, {Zhang},
  {Blake}, {Green}, {Andrews}, {Evans}, {Henning}, {{\"O}berg}, {Pontoppidan},
  {Qi}, {Salyk}, \& {van Dishoeck}}]{2013Natur.493..644B}
{Bergin}, E.~A., {Cleeves}, L.~I., {Gorti}, U., {et~al.} 2013, \nat, 493, 644

\bibitem[{{Boneberg} {et~al.}(2016){Boneberg}, {Pani{\'c}}, {Haworth},
  {Clarke}, \& {Min}}]{2016MNRAS.461..385B}
{Boneberg}, D.~M., {Pani{\'c}}, O., {Haworth}, T.~J., {Clarke}, C.~J., \&
  {Min}, M. 2016, \mnras, 461, 385

\bibitem[{{Bosman} {et~al.}(2018){Bosman}, {Walsh}, \& {van
  Dishoeck}}]{2018arXiv180802220B}
{Bosman}, A.~D., {Walsh}, C., \& {van Dishoeck}, E.~F. 2018, \aap, 618, A182

\bibitem[{{Boss}(2011)}]{2011ApJ...731...74B}
{Boss}, A.~P. 2011, \apj, 731, 74

\bibitem[{{Brinch} \& {Hogerheijde}(2010)}]{2010A&A...523A..25B}
{Brinch}, C., \& {Hogerheijde}, M.~R. 2010, \aap, 523, A25

\bibitem[{{Carney} {et~al.}(2018){Carney}, {Fedele}, {Hogerheijde}, {Favre},
  {Walsh}, {Bruderer}, {Miotello}, {Murillo}, {Klaassen}, {Henning}, \& {van
  Dishoeck}}]{2018A&A...614A.106C}
{Carney}, M.~T., {Fedele}, D., {Hogerheijde}, M.~R., {et~al.} 2018, \aap, 614,
  A106

\bibitem[{{Carney} {et~al.}(2019){Carney}, {Hogerheijde}, {Guzm{\'a}n},
  {Walsh}, {{\"O}berg}, {Fayolle}, {Cleeves}, {Carpenter}, \&
  {Qi}}]{2019A&A...623A.124C}
{Carney}, M.~T., {Hogerheijde}, M.~R., {Guzm{\'a}n}, V.~V., {et~al.} 2019,
  \aap, 623, A124

\bibitem[{{Durisen} {et~al.}(2007){Durisen}, {Boss}, {Mayer}, {Nelson},
  {Quinn}, \& {Rice}}]{2007prpl.conf..607D}
{Durisen}, R.~H., {Boss}, A.~P., {Mayer}, L., {et~al.} 2007, Protostars and
  Planets V, 607

\bibitem[{{Evans} {et~al.}(2015){Evans}, {Ilee}, {Boley}, {Caselli}, {Durisen},
  {Hartquist}, \& {Rawlings}}]{2015MNRAS.453.1147E}
{Evans}, M.~G., {Ilee}, J.~D., {Boley}, A.~C., {et~al.} 2015, \mnras, 453, 1147

\bibitem[{{Gaia Collaboration} {et~al.}(2018){Gaia Collaboration}, {Brown},
  {Vallenari}, {Prusti}, {de Bruijne}, {Babusiaux}, {Bailer-Jones}, {Biermann},
  {Evans}, {Eyer}, \& et~al.}]{2018A&A...616A...1G}
{Gaia Collaboration}, {Brown}, A.~G.~A., {Vallenari}, A., {et~al.} 2018, \aap,
  616, A1

\bibitem[{{Graedel} {et~al.}(1982){Graedel}, {Langer}, \&
  {Frerking}}]{1982ApJS...48..321G}
{Graedel}, T.~E., {Langer}, W.~D., \& {Frerking}, M.~A. 1982, \apjs, 48, 321

\bibitem[{{Hartmann} {et~al.}(2016){Hartmann}, {Herczeg}, \&
  {Calvet}}]{2016ARA&A..54..135H}
{Hartmann}, L., {Herczeg}, G., \& {Calvet}, N. 2016, \araa, 54, 135

\bibitem[{{Isella} {et~al.}(2007){Isella}, {Testi}, {Natta}, {Neri}, {Wilner},
  \& {Qi}}]{2007A&A...469..213I}
{Isella}, A., {Testi}, L., {Natta}, A., {et~al.} 2007, \aap, 469, 213

\bibitem[{{Isella} {et~al.}(2016){Isella}, {Guidi}, {Testi}, {Liu}, {Li}, {Li},
  {Weaver}, {Boehler}, {Carperter}, {De Gregorio-Monsalvo}, {Manara}, {Natta},
  {P{\'e}rez}, {Ricci}, {Sargent}, {Tazzari}, \&
  {Turner}}]{2016PhRvL.117y1101I}
{Isella}, A., {Guidi}, G., {Testi}, L., {et~al.} 2016, Physical Review Letters,
  117, 251101

\bibitem[{{Isella} {et~al.}(2018){Isella}, {Huang}, {Andrews}, {Dullemond},
  {Birnstiel}, {Zhang}, {Zhu}, {Guzm{\'a}n}, {P{\'e}rez}, {Bai}, {Benisty},
  {Carpenter}, {Ricci}, \& {Wilner}}]{2018ApJ...869L..49I}
{Isella}, A., {Huang}, J., {Andrews}, S.~M., {et~al.} 2018, \apjl, 869, L49

\bibitem[{{Kama} {et~al.}(2016){Kama}, {Bruderer}, {van Dishoeck},
  {Hogerheijde}, {Folsom}, {Miotello}, {Fedele}, {Belloche}, {G{\"u}sten}, \&
  {Wyrowski}}]{2016A&A...592A..83K}
{Kama}, M., {Bruderer}, S., {van Dishoeck}, E.~F., {et~al.} 2016, \aap, 592,
  A83

\bibitem[{Klapper {et~al.}(2003)Klapper, Surin, Lewen, Muller, Pak, \&
  Winnewisser}]{Klapper_2003}
Klapper, G., Surin, L., Lewen, F., {et~al.} 2003, The Astrophysical Journal,
  582, 262

\bibitem[{{Liu} {et~al.}(2018){Liu}, {Jin}, {Li}, {Isella}, \&
  {Li}}]{2018ApJ...857...87L}
{Liu}, S.-F., {Jin}, S., {Li}, S., {Isella}, A., \& {Li}, H. 2018, \apj, 857,
  87

\bibitem[{{Long} {et~al.}(2017){Long}, {Herczeg}, {Pascucci}, {Drabek-Maunder},
  {Mohanty}, {Testi}, {Apai}, {Hendler}, {Henning}, {Manara}, \&
  {Mulders}}]{2017ApJ...844...99L}
{Long}, F., {Herczeg}, G.~J., {Pascucci}, I., {et~al.} 2017, \apj, 844, 99

\bibitem[{{Loomis} {et~al.}(2018){Loomis}, {{\"O}berg}, {Andrews}, {Walsh},
  {Czekala}, {Huang}, \& {Rosenfeld}}]{2018AJ....155..182L}
{Loomis}, R.~A., {{\"O}berg}, K.~I., {Andrews}, S.~M., {et~al.} 2018, \aj, 155,
  182

\bibitem[{{Manara} {et~al.}(2018){Manara}, {Morbidelli}, \&
  {Guillot}}]{2018A&A...618L...3M}
{Manara}, C.~F., {Morbidelli}, A., \& {Guillot}, T. 2018, \aap, 618, L3

\bibitem[{{Mathews} {et~al.}(2013){Mathews}, {Klaassen}, {Juh{\'a}sz},
  {Harsono}, {Chapillon}, {van Dishoeck}, {Espada}, {de Gregorio-Monsalvo},
  {Hales}, {Hogerheijde}, {Mottram}, {Rawlings}, {Takahashi}, \&
  {Testi}}]{2013A&A...557A.132M}
{Mathews}, G.~S., {Klaassen}, P.~D., {Juh{\'a}sz}, A., {et~al.} 2013, \aap,
  557, A132

\bibitem[{{McClure} {et~al.}(2016){McClure}, {Bergin}, {Cleeves}, {van
  Dishoeck}, {Blake}, {Evans}, {Green}, {Henning}, {{\"O}berg}, {Pontoppidan},
  \& {Salyk}}]{2016ApJ...831..167M}
{McClure}, M.~K., {Bergin}, E.~A., {Cleeves}, L.~I., {et~al.} 2016, \apj, 831,
  167

\bibitem[{{Mendigut{\'{\i}}a} {et~al.}(2013){Mendigut{\'{\i}}a}, {Brittain},
  {Eiroa}, {Meeus}, {Montesinos}, {Mora}, {Muzerolle}, {Oudmaijer}, \&
  {Rigliaco}}]{2013ApJ...776...44M}
{Mendigut{\'{\i}}a}, I., {Brittain}, S., {Eiroa}, C., {et~al.} 2013, \apj, 776,
  44

\bibitem[{{Miotello} {et~al.}(2014){Miotello}, {Bruderer}, \& {van
  Dishoeck}}]{2014A&A...572A..96M}
{Miotello}, A., {Bruderer}, S., \& {van Dishoeck}, E.~F. 2014, \aap, 572, A96

\bibitem[{{Molyarova} {et~al.}(2017){Molyarova}, {Akimkin}, {Semenov},
  {Henning}, {Vasyunin}, \& {Wiebe}}]{2017ApJ...849..130M}
{Molyarova}, T., {Akimkin}, V., {Semenov}, D., {et~al.} 2017, \apj, 849, 130

\bibitem[{{M{\"u}ller} {et~al.}(2005){M{\"u}ller}, {Schl{\"o}der}, {Stutzki},
  \& {Winnewisser}}]{2005JMoSt.742..215M}
{M{\"u}ller}, H.~S.~P., {Schl{\"o}der}, F., {Stutzki}, J., \& {Winnewisser}, G.
  2005, JMoSt, 742, 215

\bibitem[{{Notsu} {et~al.}(2019){Notsu}, {Akiyama}, {Booth}, {Nomura}, {Walsh},
  {Hirota}, {Honda}, {Tsukagoshi}, \& {Millar}}]{2019ApJ...875...96N}
{Notsu}, S., {Akiyama}, E., {Booth}, A., {et~al.} 2019, \apj, 875, 96

\bibitem[{{Pinte} {et~al.}(2018){Pinte}, {Price}, {M{\'e}nard}, {Duch{\^e}ne},
  {Dent}, {Hill}, {de Gregorio-Monsalvo}, {Hales}, \&
  {Mentiplay}}]{2018ApJ...860L..13P}
{Pinte}, C., {Price}, D.~J., {M{\'e}nard}, F., {et~al.} 2018, \apjl, 860, L13

\bibitem[{{Powell} {et~al.}(2019){Powell}, {Murray-Clay}, {P{\'e}rez},
  {Schlichting}, \& {Rosenthal}}]{2019ApJ...878..116P}
{Powell}, D., {Murray-Clay}, R., {P{\'e}rez}, L.~M., {Schlichting}, H.~E., \&
  {Rosenthal}, M. 2019, \apj, 878, 116

\bibitem[{{Qi} {et~al.}(2011){Qi}, {D'Alessio}, {{\"O}berg}, {Wilner},
  {Hughes}, {Andrews}, \& {Ayala}}]{2011ApJ...740...84Q}
{Qi}, C., {D'Alessio}, P., {{\"O}berg}, K.~I., {et~al.} 2011, \apj, 740, 84

\bibitem[{{Qi} {et~al.}(2015){Qi}, {{\"O}berg}, {Andrews}, {Wilner}, {Bergin},
  {Hughes}, {Hogherheijde}, \& {D'Alessio}}]{2015ApJ...813..128Q}
{Qi}, C., {{\"O}berg}, K.~I., {Andrews}, S.~M., {et~al.} 2015, \apj, 813, 128

\bibitem[{{Rice} {et~al.}(2004){Rice}, {Lodato}, {Pringle}, {Armitage}, \&
  {Bonnell}}]{2004MNRAS.355..543R}
{Rice}, W.~K.~M., {Lodato}, G., {Pringle}, J.~E., {Armitage}, P.~J., \&
  {Bonnell}, I.~A. 2004, \mnras, 355, 543

\bibitem[{{Teague} {et~al.}(2018){Teague}, {Bae}, {Bergin}, {Birnstiel}, \&
  {Foreman-Mackey}}]{2018ApJ...860L..12T}
{Teague}, R., {Bae}, J., {Bergin}, E.~A., {Birnstiel}, T., \& {Foreman-Mackey},
  D. 2018, \apjl, 860, L12

\bibitem[{{Tilling} {et~al.}(2012){Tilling}, {Woitke}, {Meeus}, {Mora},
  {Montesinos}, {Riviere-Marichalar}, {Eiroa}, {Thi}, {Isella}, {Roberge},
  {Martin-Zaidi}, {Kamp}, {Pinte}, {Sandell}, {Vacca}, {M{\'e}nard},
  {Mendigut{\'{\i}}a}, {Duch{\^e}ne}, {Dent}, {Aresu}, {Meijerink}, \&
  {Spaans}}]{2012A&A...538A..20T}
{Tilling}, I., {Woitke}, P., {Meeus}, G., {et~al.} 2012, \aap, 538, A20

\bibitem[{{Toomre}(1964)}]{1964ApJ...139.1217T}
{Toomre}, A. 1964, \apj, 139, 1217

\bibitem[{{van den Ancker} {et~al.}(1998){van den Ancker}, {de Winter}, \&
  {Tjin A Djie}}]{1998A&A...330..145V}
{van den Ancker}, M.~E., {de Winter}, D., \& {Tjin A Djie}, H.~R.~E. 1998,
  \aap, 330, 145

\bibitem[{{Venuti} {et~al.}(2014){Venuti}, {Bouvier}, {Flaccomio}, {Alencar},
  {Irwin}, {Stauffer}, {Cody}, {Teixeira}, {Sousa}, {Micela}, {Cuillandre}, \&
  {Peres}}]{2014A&A...570A..82V}
{Venuti}, L., {Bouvier}, J., {Flaccomio}, E., {et~al.} 2014, \aap, 570, A82

\bibitem[{{Walsh} {et~al.}(2010){Walsh}, {Millar}, \&
  {Nomura}}]{2010ApJ...722.1607W}
{Walsh}, C., {Millar}, T.~J., \& {Nomura}, H. 2010, \apj, 722, 1607

\bibitem[{{Williams} \& {McPartland}(2016)}]{2016ApJ...830...32W}
{Williams}, J.~P., \& {McPartland}, C. 2016, \apj, 830, 32

\bibitem[{{Williams J.~P. and Best W.~M.~J.}(2014)}]{2014ApJ...788...59W}
{Williams J.~P. and Best W.~M.~J.} 2014, \apj, 788, 59

\bibitem[{{Wilson}(1999)}]{0034-4885-62-2-002}
{Wilson}, T.~L. 1999, Reports on Progress in Physics, 62, 143

\bibitem[{{Woitke} {et~al.}(2019){Woitke}, {Kamp}, {Antonellini}, {Anthonioz},
  {Baldovin-Saveedra}, {Carmona}, {Dionatos}, {Dominik}, {Greaves},
  {G{\"u}del}, {Ilee}, {Liebhardt}, {Menard}, {Min}, {Pinte}, {Rab}, {Rigon},
  {Thi}, {Thureau}, \& {Waters}}]{2019PASP..131f4301W}
{Woitke}, P., {Kamp}, I., {Antonellini}, S., {et~al.} 2019, \pasp, 131, 064301

\bibitem[{{Zhang} {et~al.}(2017){Zhang}, {Bergin}, {Blake}, {Cleeves}, \&
  {Schwarz}}]{2017NatAs...1E.130Z}
{Zhang}, K., {Bergin}, E.~A., {Blake}, G.~A., {Cleeves}, L.~I., \& {Schwarz},
  K.~R. 2017, Nature Astronomy, 1, 0130

\bibitem[{{Zhu} {et~al.}(2019){Zhu}, {Zhang}, {Jiang}, {Kataoka}, {Birnstiel},
  {Dullemond}, {Andrews}, {Huang}, {P{\'e}rez}, {Carpenter}, {Bai}, {Wilner},
  \& {Ricci}}]{2019arXiv190402127Z}
{Zhu}, Z., {Zhang}, S., {Jiang}, Y.-F., {et~al.} 2019, \apjl, 877, L18

\end{thebibliography}

\end{document}